\documentclass[
    a4paper,
    notitlepage,
    10pt,
]{article}

\usepackage[utf8]{inputenc}
\usepackage[LGR,T1]{fontenc}
\usepackage[british]{babel}

\usepackage{lmodern}
\usepackage{amsmath}
\usepackage[margin=2cm]{geometry}
\usepackage{hyperref}
\usepackage{graphicx}
\usepackage{amsfonts}
\usepackage{caption}
\usepackage{siunitx}
\usepackage{adjustbox}
\usepackage{subcaption}
\usepackage[babel=true]{microtype}
\usepackage{parskip}
\usepackage{abstract}
\usepackage{hyphenat}
\usepackage{mhchem}
\usepackage{authblk}
\usepackage{xcolor}

\usepackage[
    style=ieee,
    citestyle=numeric-comp,
    sorting=none,
    sortcites,
    giveninits,
    maxnames=5,
    backend=biber,
    urldate=iso,
    date=year,
]{biblatex}
\usepackage{csquotes}

\definecolor{CoolBlack}{rgb}{0.0, 0.18, 0.39}
\hypersetup{
    hidelinks,
    colorlinks,
    linkcolor=blue,
    filecolor=magenta,
    urlcolor=cyan,
    citecolor=CoolBlack,
}
\DeclareSIUnit{\octave}{oct}

\title{Millisecond-resolved infrared spectroscopy study of polymer brush swelling dynamics}

\author[1,*]{K.F.A. Jorissen}
\author[2]{L.B. Veldscholte}
\author[1]{M. Odijk}
\author[2]{S. de Beer}
\affil[1]{BIOS Lab on a Chip Group, EEMCS, MESA+ Institute, University of Twente, the Netherlands}
\affil[2]{Functional Polymer Surfaces, Department of Molecules and Materials, MESA+ Institute, University of Twente, the Netherlands}
\affil[*]{k.f.a.jorissen@utwente.nl}

\begin{document}
\twocolumn[\maketitle
%\centering\includegraphics{img/TOC.pdf}
\begin{abstract}
	\vspace{2em}
    We present the study of millisecond-resolved polymer brush swelling dynamics using infrared spectroscopy with a custom-built quantum cascade laser-based infrared spectrometer at a \SI{1}{\kilo\hertz} sampling rate after averaging. By cycling the humidity of the environment of the polymer brush, we are able to measure the swelling dynamics sequentially at different wavenumbers. The high sampling rate provides us with information on the reconformation of the brush at a higher temporal resolution than previously reported. Using spectroscopic ellipsometry, we study the brush swelling dynamics as a reference experiment and to correct artefacts of the infrared measurement approach. This technique informs on the changes in the brush thickness and refractive index. Our results indicate that the swelling dynamics of the polymer brush are poorly described by Fickian diffusion and the Berens-Hopfenberg formalism, pointing toward more complicated underlying transport.  
	\vspace{2em}
\end{abstract}
]

\section{Introduction}
% IR spectroscopy
Time-resolved infrared (TR-IR) spectroscopy is used to study the transient behaviour of IR-active bonds. On the micro- and nanosecond timescale, transient behaviour informs on the transport and kinetic behaviour of chemical systems \cite{cuestaATRSEIRASTimeresolvedStudies2022}. TR-IR spectroscopy has traditionally been used in Fourier transform (FT) IR systems, with which temporal resolutions of \SI{100}{\mega\hertz} can be achieved in step-scan mode \cite{uhmannTimeResolvedFTIRAbsorption1991, johnsonApplicationsTimeResolvedStepScan1993}. Recent advancements in quantum cascade laser (QCL) technology \cite{faistQuantumCascadeLaser1994, razeghiQuantumCascadeLasers2015, hugiExternalCavityQuantum2010a} have enabled higher temporal resolution sampling at discrete frequencies by providing a widely tunable, high-intensity light source \cite{childsSensitivityAdvantageQCL2015, dabrowskaMidIRDispersionSpectroscopy2023}. Moreover, QCL frequency combs have enabled the acquisition of full spectra at nanosecond timescales \cite{villaresDualcombSpectroscopyBased2014} and are now commercially available \cite{geiserQCLDualCombSpectroscopy2021}. However, with all approaches, the signal-to-noise ratio deteriorates for higher temporal resolutions, forming a trade-off between spectral and temporal resolution. Using discrete\hyp{}frequency infrared (DF-IR) spectroscopy, the attainable signal-to-noise ratio is higher than full spectral approaches due to a higher signal intensity at single wavelengths \cite{phalDesignConsiderationsDiscrete2021}. Thus, for single transient signals, DF-IR can provide a higher time resolution than approaches where a full spectrum is attained. Additionally, for highly repeatable systems, DF-IR can be used iteratively to acquire time-resolved information at separate wavelengths. 
% Microfluidics
When studying small systems using infrared spectroscopy, controlling the environment of the studied system can be challenging. One method is to place the studied system in a microfluidic environment. Microfluidics and infrared spectroscopy can be integrated through different approaches, such as thin film waveguides \cite{schadleMidInfraredWaveguidesPerspective2016, wangUltrasensitiveMidinfraredEvanescent2012}, reflective measurements, transmission measurements \cite{andrewchanRapidPrototypingMicrofluidic2010, srisa-artIRCompatiblePDMSMicrofluidic2018} and attenuated total reflection (ATR) \cite{perroCombiningMicrofluidicsFTIR2016, chanChemicalImagingMicrofluidic2009}. In microfluidics, the high surface-to-volume ratio leads to a larger influence of surface effects. Consequently, surface-sensitive techniques such as waveguides, infrared reflection absorption spectroscopy \cite{hollinsInfraredReflectionAbsorption2006}, and ATR spectroscopy are often used. 

QCL-based DF-IR spectroscopy can be used to elucidate the dynamic chemical behaviour of systems on a smaller timescale than previously reported. A relevant subject to study using this principle is the water sorption dynamics in polymer films or brushes.
% Swelling dynamics in polymer brushes
Polymer brushes are coatings consisting of macromolecules densely grafted to a substrate. At sufficiently high grafting densities, they are forced to stretch away from the substrate, forming a `brush' structure, with the chains extending away from the substrate \cite{milner_polymer_1991, zhao_polymer_2000}. Polymer brushes exhibit properties that can be considerably different from non-grafted films or bulk polymers \cite{brittain_structural_2007,Bhayo2022}, which gives rise to a broad range of applications in the (bio-)medical field~\cite{Li2021_review,Nastyshyn2022} in anti-fouling surfaces \cite{higakiAntifoulingBehaviorPolymer2016,Maan2020}, lubricious surfaces \cite{kreerPolymerbrushLubricationReview2016,Yan2019}, sensors~\cite{Sanchez2021} and separation membranes \cite{keatingPolymerBrushesMembrane2016,Durmaz2021}. In several of these applications, brushes are swollen by vapours \cite{Ritsema2022review}, such as water \cite{Galvin2014,ritsema_van_eck_vapor_2022}. Understanding brush swelling dynamics is imperative for proper application of brushes in these contexts.
% Swelling dynamics / diffusion in polymers
Although time-resolved brush vapour swelling has been preliminarily reported \cite{horst_swelling_2020, kap_nonequilibrium_2023}, the dynamics of vapour sorption have not been extensively explored. Vapour sorption dynamics is still an active area of research in non-grafted polymer films \cite{ogiegloSituEllipsometryStudies2015, tempelman_swelling_2019}, and grafting and the resulting anisotropic chain conformation in brushes further complicate solvent diffusion. Polymer films are typically glassy, which complicates the swelling dynamics, as they can undergo a solvent-induced glass transition during swelling, impacting the diffusivity. A glassy polymer can also contain voids, which the solvent can occupy without diluting the film \cite{tempelman_swelling_2019}.
Moreover, variable brush thickness, changes in the polymer matrix affecting diffusivity, and the potential gradient arising from solvent-polymer interactions lead to anomalous (or non-Fickian) diffusion. The interplay between solvent diffusion, polymer-solvent interaction energy and polymer conformation kinetics make descriptive swelling kinetic models complex. \cite{papanuTransportModelsSwelling1989}

In this work, a custom-built QCL-based DF-IR spectrometer is used to study the swelling dynamics of a polymer brush in an ATR configuration. The ATR crystal coated with polymer brush is enclosed in a microfluidic flow cell to accurately control its environment. The spectrometer is configured to measure polymer brush swelling dynamics at a high time resolution while still capturing a full spectrum. By integrating high-speed infrared spectroscopy with rapid humidity switching, we are able to show swelling kinetics at a higher time resolution than previously reported (to our knowledge). The infrared spectroscopy is complimented by in-situ visible light spectroscopic ellipsometry to measure the swelling and refractive index change.

\begin{figure}[bt]
    \centering
    \includegraphics[width=\linewidth]{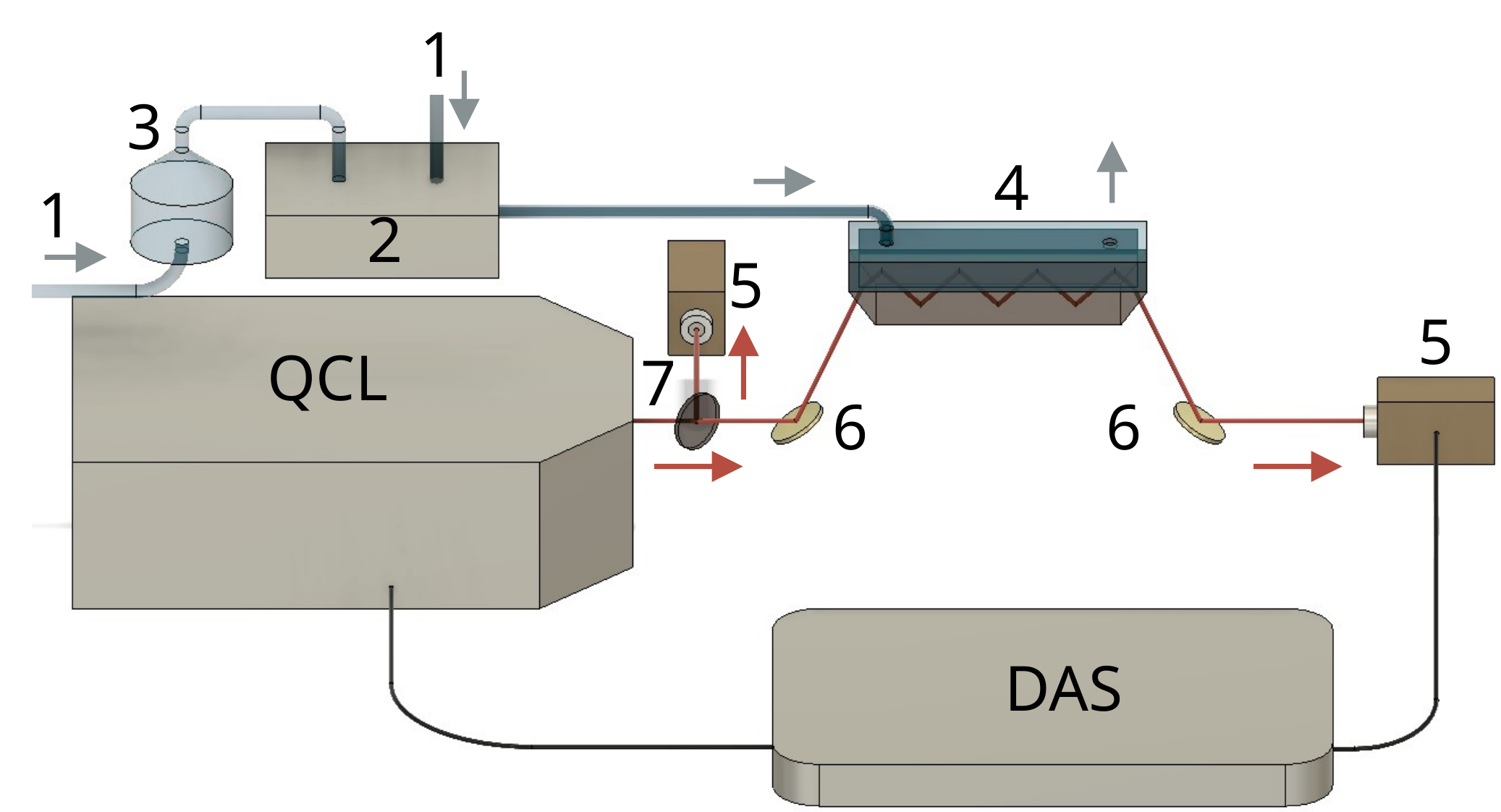}
    \caption{A schematic depiction of the infrared spectrometer. Red arrows indicate light direction. Blue arrows indicate airflow direction. QCL: Quantum cascade laser light source. DAS: Field programmable gate array based data acquisition system.  1. Dry nitrogen source. 2. OpenHumidistat humidity controller with added solenoid valve system. 3. Bubbler. 4.   Silicon attenuated total reflection flow cell, coated with polymer brush and enclosed by thermoformed polymer enclosure. 5. Mercury cadmium telluride infrared detectors. 6. Gold reflective mirrors. 7. Zinc selenide beam splitter.}
    \label{fig:IRspec_schematic}
\end{figure}

\begin{figure}[bt]
    \centering
    \includegraphics[width=\linewidth]{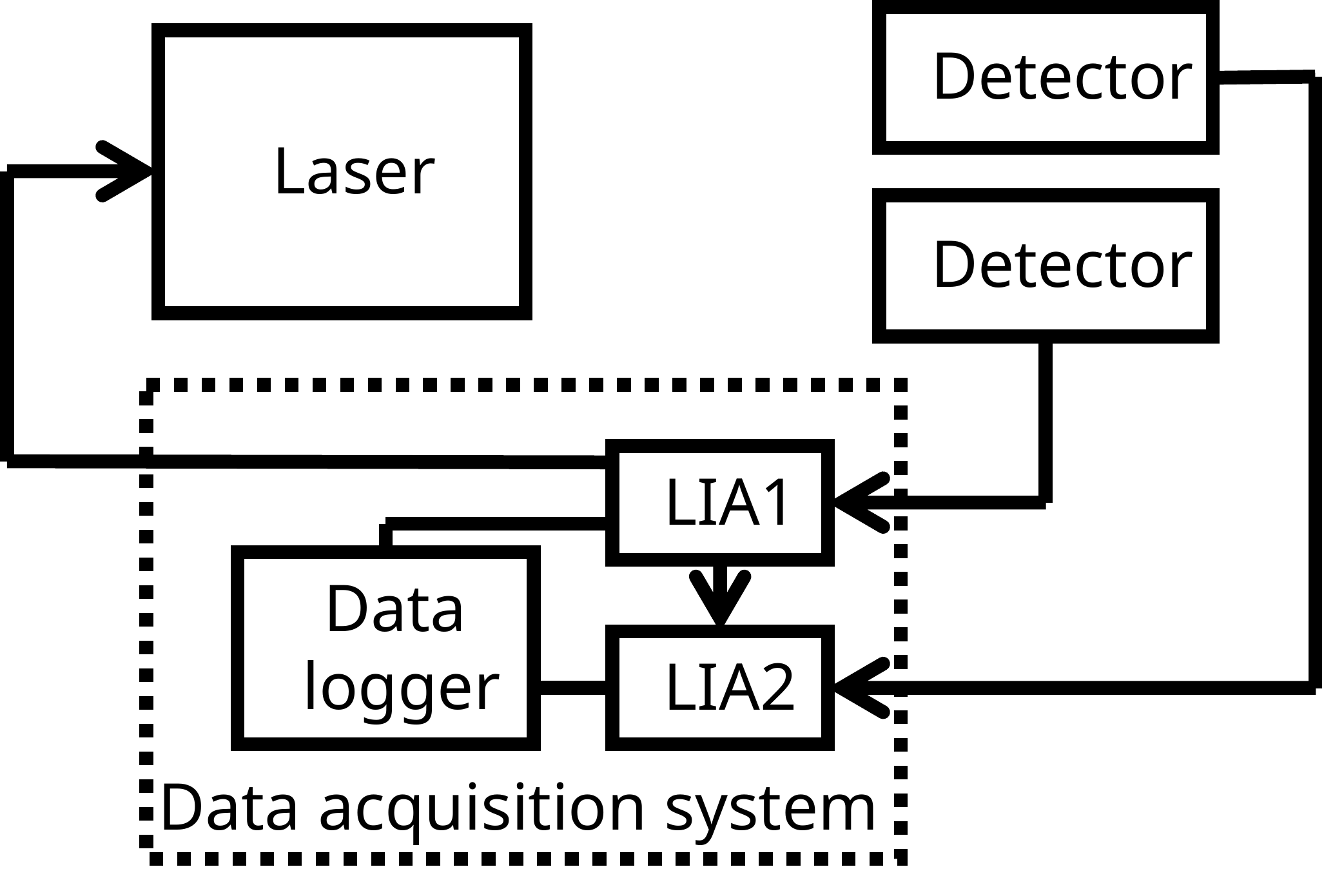}
    \caption{A schematic depicting the logic flow of the data acquisition system; the reference signal of lock-in amplifier (LIA) 1 drives the laser and is also used as the reference signal of lock-in amplifier 2. The detector signal is processed by the lock-in amplifiers and logged in the data logger module.}
    \label{fig:IRspec_DAS}
\end{figure}

\section{Methods}
\subsection{Brush synthesis}
A poly(3-sulfopropyl methacrylate) brush is prepared in a grafting-from manner on silicon substrates by an oxygen-tolerant, surface-initiated controlled radical polymerisation known as activators regenerated by electron transfer atom transfer radical polymerisation (SI-ARGET-ATRP). The substrates are first cleaned and activated by oxygen plasma and subsequently functionalised with a silane anchor (3-amino\-propyl\-tri\-ethoxy\-silane) by vapour deposition. An initiator (2-bromoisobutyryl bromide) is then coupled. Finally, the polymerisation is conducted from the initiator\hyp{}functionalised substrates for \SI{2}{\hour}, resulting in brushes with a dry height of approximately \SI{175}{\nano\meter}. This synthesis procedure is described in detail by \textcite{veldscholteScalableAirTolerantMLVolume2023}.

For IR spectroscopy, the brushes are grown on silicon ATR crystals. For ellipsometry, the substrates are 5x2 \si{\centi\meter} pieces of a silicon wafer (Okmetic, Finland) \SI{525}{\micro\meter} thick, boron-doped with a (100) crystal lattice orientation).

\subsection{Infrared spectrometer design}
The global design of the infrared spectrometer is shown in \autoref{fig:IRspec_schematic}. The data processing and triggering schematic is shown in \autoref{fig:IRspec_DAS}. The infrared spectrometer is built up from an external cavity quantum cascade laser (two separate internal gain media, Mircat-QT, Daylight solutions, US), beam path including sample, mercury cadmium telluride (MCT) detector (PVI-4TE-10.6, Vigo Photonics, Poland) with pre-amplifier (PIP-UC-HS, Vigo Photonics, Poland) and thermoelectric controller (PTCC-01-BAS, Vigo Photonics, Poland), and an FPGA-based data acquisition system (Moku: Pro, Liquid Instruments, Australia).

The QCL has a frequency range of \SIrange{1187}{1857}{\per\centi\meter}. The QCL has a minimum step size of \SI{0.5}{\per\centi\meter} and a $\leq$ \SI{1}{\per\centi\meter} full-width-half-maximum wavelength accuracy. The QCL is operated in an externally driven pulsed mode, pulsing at a frequency defined in the data acquisition system. The pulse width is independently set in the QCL, and can be set up to a duty cycle of 30\%. The QCL is programmed to step from wavenumber to wavenumber at a fixed time interval, sweeping over the entire wavelength range in a single full experiment.

The MCT detectors used are identical. The first detector is the data acquisition detector, whereas the other detector is used to identify wavelength steps in data processing. The MCT detector pre-amplifiers are tuned identically before each experiment to maximise the signal-to-noise ratio by changing the bias voltage.

The data acquisition system contains four 600 MHz analog-to-digital converters integrated with four internal field programmable arrays (FPGAs), of which three are used: two as a lock-in amplifier and one as a data logger. This is shown in \autoref{fig:IRspec_DAS}. Data acquisition of the pulsed signal from the detectors is processed using the two lock-in amplifiers. The first lock-in amplifier generates a reference clock which is also used as a reference signal in the second, ensuring the two lock-in amplifiers do not drift with respect to each other. The clock signal in the first lock-in amplifier is also amplified and used to drive the QCL's TTL pulse port. Before each experiment, the phase of the lock-in amplifier is tuned to maximise the signal-to-noise ratio, and only the in-phase element is logged (as opposed to the quadrature). The built-in data logger is limited to \SI{2}{\mega\hertz} logging, so each data stream gets logged at \SI{1}{\mega\hertz}. An inbuilt averager can be used to average data while logging when data averaging is anticipated. By using an FPGA multi-instrument system the spectrometer data processing can be reconfigured without altering hardware, depending on measurement needs. 

The beam path is controlled by a zinc selenide beam-splitter, two gold mirrors, and the silicon ATR crystal. The fabrication of the silicon ATR crystal is reported by \textcite{lozemanModularMicroreactorIntegrated2020}. The zinc selenide beam splitter allows separate logging of the laser signal intensity and the signal intensity after passing through the ATR crystal, both of which are used in data processing. The alignment setup of the crystal through the gold mirrors is made in-house. The optical alignment of the mirrors is optimised before each measurement. Notably, a wavenumber-dependent pointing error in the QCL affects the alignment. Because of this, the alignment is tuned for both high and low wavenumbers within the spectral range of the laser. Nonetheless, a pointing error induced alignment variation results in reduced signal quality at the extremes of the spectral range. 

\subsection{IR spectroscopy}
The infrared spectrometer is set to sweep from \SIrange{1187}{1857}{\per\centi\meter} in steps of \SI{0.5}{\per\centi\meter}. The first lock-in amplifier is set at \SI{2}{\mega\hertz}, combined with a \SI{150}{\nano\second} peak width for the QCL. Both lock-in amplifiers output the in-phase component of the mixed signal after a \SI{200}{\hertz}, \SI{-18}{\decibel\per\octave} low-pass filter. Data is logged at \SI{1}{\kilo\hertz}, averaging the signal between data points.

To control the environment of the brush on the ATR crystal, it is bonded with a cap and turned into a flow cell. The cap is made using 2mm thick styrene-ethylene-butylene-styrene elastomer (Flexdym, Eden Microfluidics, France) thermoformed on a polytetrafluoroethylene mould at \SI{140}{\celsius}. The mould is \SI{0.7}{\milli\meter} high and has an area of \SI{193}{\milli\meter\squared}. An inlet and an outlet hole are punched into the cap after forming. The cap is bonded to the brush-covered ATR crystal in an oven at \SI{80}{\celsius} while pressed. The resulting flow cell is connected to the humidistat-valve humidity control system using adhesive connectors and flexible polymer tubing.

\subsection{Humidity control}
To perform humidity-controlled experiments, we use our OpenHumidistat \cite{veldscholte_openhumidistat_2022} humidity controller, which provides closed-loop control over the humidity of a nitrogen stream at a constant flow of \SI{2}{\liter\per\minute}. As the humidity sensor for feedback to the humidity controller cannot fit inside either the microfluidic flow cell (for IR spectroscopy) or the measurement chamber (for ellipsometry), we use a small pre-chamber containing the humidity sensor.

Because we are interested in the dynamic swelling response of our samples on a short time scale, we only use the humidity controller to provide a nitrogen stream of constant, stable 90\% humidity. We use a solenoid valve system (Festo, The Netherlands) to quickly select between humidified and dry nitrogen to feed into the IR spectroscopy and ellipsometry flow cells. Using this setup, we can impose a rapid change in the humidity without being limited by the transient response of the humidistat.

The internal volume of the flow chamber is \SI{135}{\micro\liter} in the infrared spectrometer. The controlled flow of \SI{2}{\liter\per\minute} results in an average residence time of \SI{4.05}{\milli\second}. The solenoid valve system switches the humidity in sync with the wavelength triggers of the QCL. Each distinct wavenumber is maintained for \SI{20}{\second}: \SI{10}{\second} for high and \SI{10}{\second} for low humidity.

For the IR spectroscopy, the synchronisation between the humidity control system and the spectrometer is arranged using transistor-transistor logic (TTL). Upon a wave\-length change, the QCL outputs a TTL pulse. A microcontroller (Arduino Uno, Arduino, Italy) is used to listen to this pulse and trigger a Matlab script on a laptop. The script, subsequently, triggers a switch in the valve system leading the desired gas feed (humidified or dry nitrogen) into the microfluidic flow chamber.

To feed the humidity-controlled nitrogen over the brush in a controlled fashion, a flow chamber around the brush is created by bonding a cap to the ATR crystal. The outlet of the valve system is connected to the flow chamber inlet, whereas the outlet does not feature tubing, but exhausts directly into the environment.

For the time coordination between the ellipsometer and humidity control system, a Matlab script is used to set the duration of the high humidity window. The script is started manually at the same time as the ellipsometry measurement. In the ellipsometer, a compatible flow cell (Heated Liquid Cell, J.A. Woollam, US) with an internal volume of 5 mL is mounted on the silicon chip, controlling the brush environment. The cell's heating feature is not used.

\subsection{Ellipsometry}
In-situ ellipsometry is conducted using a spectroscopic ellipsometer (M-2000X, J.A. Woollam, US). The measurements are performed in in-situ mode at wavelengths between \SI{350} and \SI{1000}{\nano\meter} at an angle of incidence of \SI{70}{\degree}. During a measurement, the humidity of the air is switched from dry to a higher humidity level, and the transient swelling response of the brush is measured. 

The data is fitted to a model composed of a Cauchy layer on top of a Si substrate. The thickness and Cauchy A and B coefficients are fitted. We do not use higher-order Cauchy coefficients, and we assume the polymer is optically transparent over the measured wavelength range.

\subsection{Data processing}
\begin{figure}[bt]
    \centering
    \includegraphics[width=\linewidth]{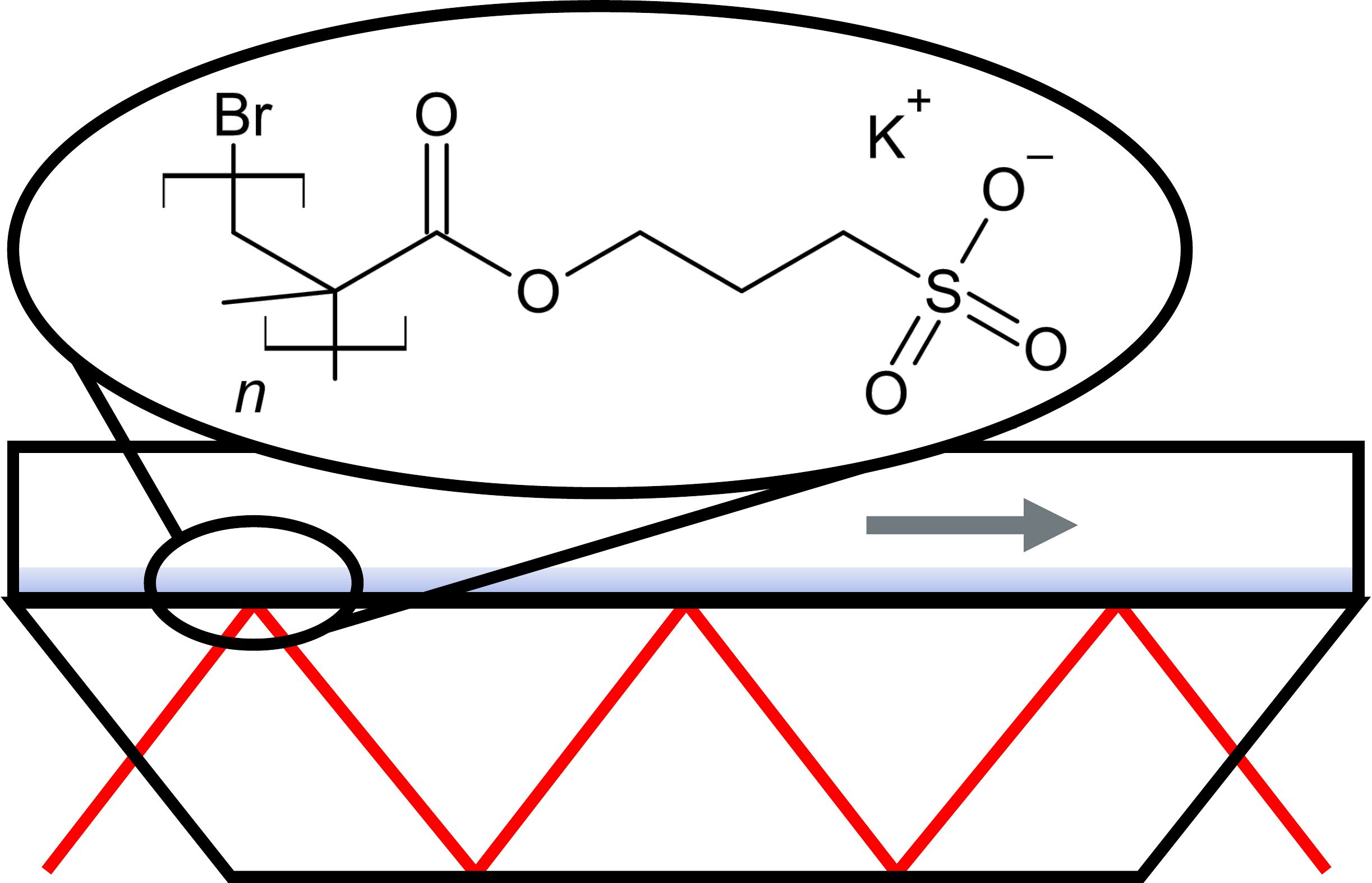}
    \caption{A schematic of the flow cell measurement setup, not to scale. A poly(SPMA) brush is grafted from the ATR crystal, and enclosed by a thermoformed polymer cap, forming a flow cell. }
    \label{fig:Cell_graphic}
\end{figure}
\label{sec:DP}
Infrared spectroscopy data processing is done in Python, the code for which can be found in SI-3. To process the infrared spectroscopy data, the twenty-second single-frequency time periods are cut into bins. Due to slight time variations in the QCL, not each twenty-second frame is equally long. Considering a small step in the wavenumber typically leads to a significant step in the laser output power, the laser power can be used to identify a wavenumber step. To identify the wavenumber steps on the relevant timescale of the measurement, the secondary detector bypassing the attenuation crystal is used to measure steps in the laser signal intensity. A peak-finding algorithm is used on the derivative in time of the secondary column signal to identify steps in the wavenumber intensity. Any bins deviating by more than 1\% of the median bin length are discarded.

The infrared absorption of the brush in humid nitrogen is calculated with respect to the dry brush. The dry brush intensity is chosen within a frame where no drying effects take place (the average signal of 7.5 to 9 seconds in the drying phase is taken), where the absorption profile is constant. The absorbance is given by \autoref{eq:abs}. Here, $I_{\rm R}$ and $I_{\rm S}$ are the intensity of the reference period and of the time-resolved signal, respectively. 
\begin{equation}
    A = \log_{10} \left(\frac{I_{\rm R}}{I_{\rm S}}\right)
    \label{eq:abs}
\end{equation}

\begin{figure}[bt]
    \centering
    \includegraphics{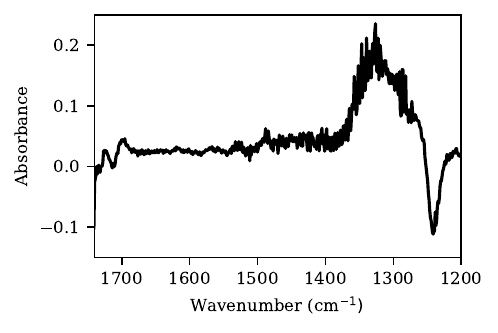}
    \caption{The absorption of the brush, half a second after being exposed to nitrogen at 90\% humidity, compared to the brush in dry nitrogen. }
    \label{fig:halfsecond}
\end{figure}

\begin{figure*}[bt]
    \centering
    \includegraphics[width=\textwidth]{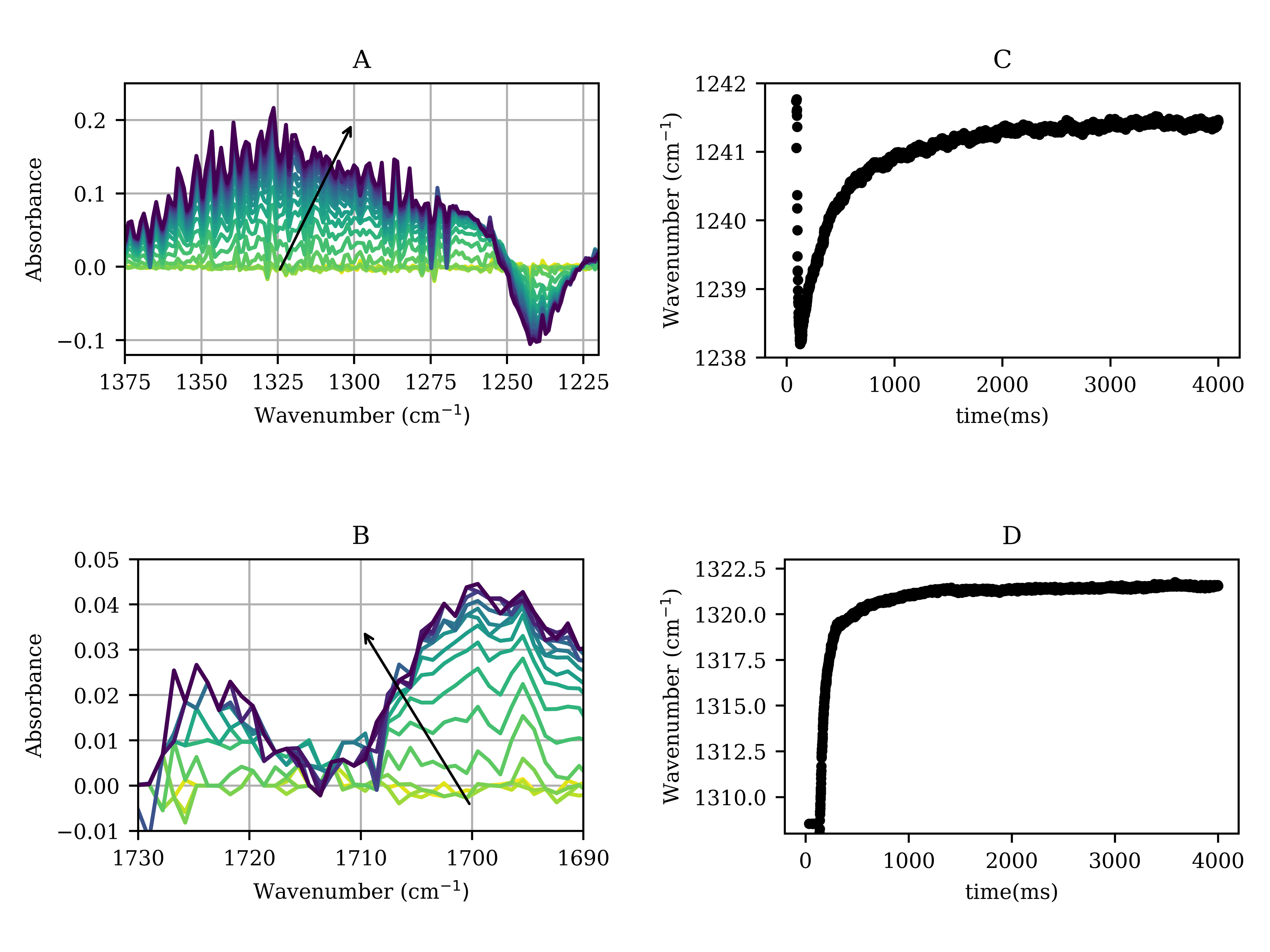}
    \caption{The absorption profiles of the polymer brush swelling in time. A. The absorption peaks at \SI{1320}{\per\centi\meter} and \SI{1242}{\per\centi\meter} growing in as after a humidity switch, in \SI{20}{\milli\second} increments up to \SI{400}{\milli\second}. The arrow indicates ascending time. B. The absorption peaks at \SI{1700}{\per\centi\meter} and \SI{1720}{\per\centi\meter} growing in as after a humidity switch, in \SI{30}{\milli\second} increments up to \SI{400}{\milli\second}. The arrow indicates ascending time. C. A time trace of the peak position of the \SI{1242}{\per\centi\meter} peak at \SI{1}{\milli\second} resolution, determined by fitting a Gaussian peak. D. A time trace of the peak position of the \SI{1320}{\per\centi\meter} peak at \SI{1}{\milli\second} resolution, determined by fitting a Gaussian peak.}
    \label{fig:multiplot}
\end{figure*}

Due to the ATR measurement approach, corrections can be made based on varying parameters affecting the effective path length through the brush: a correction for the evanescent wave thickness varying with the refractive index and wavelength, and a correction for the overlap of the evanescent wave and the brush thickness. The evanescent wave thickness in ATR is given by \autoref{eq:dp}, where $|z|$ is the distance from the surface where light intensity has decayed 1/e, $\lambda$ is the wavelength, $\theta$ is the incident angle, and $n$ is the ratio of the refractive indices of the sample and the crystal. \cite{pedrotti_introduction_2017}. Due to the variation of the refractive index of the brush as a function of the swelling, the thickness of the sensing window varies. Additionally, the sensing window depth is dependent on the wavelength. Finally, the sensing window is limited by the brush thickness. Thus, any absorption found in the polymer brush is not only molecular but also a function of these measurement artefacts.
\begin{equation}
    |z| = \frac{\lambda}{2\pi\sqrt{\frac{\sin^2(\theta)}{n^2}}-1}
\label{eq:dp}
\end{equation}
The dimensionless brush thickness overlap with the evanescent wave model is given by \autoref{eq:overlap}. Here, $d$ is the thickness of the brush.
\begin{equation}
    O = \frac{|z|}{|z|-|z|e^{-\frac{d}{|z|}}}
    \label{eq:overlap}
\end{equation}
Combining \autoref{eq:dp} and \ref{eq:overlap}, $O$ functions as a dimensionless correction factor between 0 and 1 for the aforementioned ATR artefacts, which is applied. While the ellipsometry data is taken at a different timescale and in the visible light regime, the swelling and refractive index information from it can be used to correct the ATR artefacts. Linear interpolation is used for ellipsometry between data points, and used to inform the infrared spectroscopy correction.
\section{Results}

\subsection{Infrared spectroscopy}
The absorption spectrum of the humid brush is shown in \autoref{fig:halfsecond}. The chemical structure of poly(SPMA) can be seen in the inset of \autoref{fig:Cell_graphic}. Note that all IR spectra shown are relative to the brush in the dry state, and therefore only show peaks changing in absorption with humidity, as opposed to all absorbing peaks. This is chosen because the alignment of the crystal is not fully reproducible. Realigning between a reference and signal measurement would introduce alignment artifacts in the absorbance signal. Additionally, this way the time between the reference and absorption measurement is smaller, minimising laser or sensor drift effects on the measurement. We focus on three peaks: the negative peak at \SI{1242}{\per\centi\meter} is attributed to be a \ce{C-O} ester stretch and the peak at \SI{1725}{\per\centi\meter} is characteristic of the ester carbonyl (\ce{C=O}) stretch. The \SI{1320}{\per\centi\meter} peak corresponds to a sulfonate S=O stretch in anhydrous alkyl sulfonic acids. However, the peak grows in the positive direction (increased absorbance) with increasing humidity, as opposed to an expected negative band. Time-resolved IR spectra in steps of \SI{20}{\milli\second} of these three peaks are shown in \autoref{fig:multiplot} A and B. An animated plot of the developing spectrum in a higher temporal resolution can be found in SI-2. Interestingly, the \SI{1320}{\per\centi\meter} bond develops as two peaks initially, one at \SI{1258}{\per\centi\meter} and one at \SI{1343}{\per\centi\meter}. Throughout the spectrum, a series of sharp peaks can be seen on top of the ingrowing peaks at a spacing of \SI{3}{\per\centi\meter}. These peaks are constant in time, and grow in with the underlying spectrum. This is characteristic for a signal going through a limited bandwidth medium, such as the ATR crystal. \cite{lin-vienHandbookInfraredRaman1991a, socratesInfraredRamanCharacteristic2004}

Tracing the positions in time by fitting a Gaussian curve to the \SI{1320}{\per\centi\meter} and \SI{1242}{\per\centi\meter} peaks and logging the position, it can be seen in \autoref{fig:multiplot} C and D that both peaks increase in wavenumber as humidity increases. Additionally, the \SI{1320}{\per\centi\meter} peak reaches equilibrium quicker than the \SI{1242}{\per\centi\meter} peak. Blueshift upon hydrogen bonding is typical for stretching modes as opposed to bending modes which typically red shift upon hydrogen bond formation\cite{susi17StrengthHydrogen1972}, pointing towards bending modes for both the \SI{1242}{\per\centi\meter} and the \SI{1320}{\per\centi\meter} peaks. 

\subsection{Ellipsometry}
\begin{figure}[tb]
    \centering
    \includegraphics{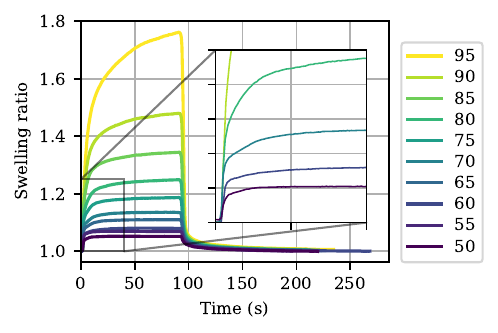}
    \caption{Time-resolved swelling curves for a range of humidity values, as measured by in-situ ellipsometry. The brushes start dry and are exposed to humidity at $t=0$, and are dried again from $t=\SI{90}{\second}$.\label{fig:elip}}
\end{figure}

\autoref{fig:elip} shows the results of the swelling experiments as measured by in-situ ellipsometry. The brushes are initially exposed to dry nitrogen and allowed to dry. At $t=0$, the solenoid valve system switches the flow through the measurement cell to humidified nitrogen. After \SI{90}{\second}, the flow is switched back to dry nitrogen. Meanwhile, the brush thickness is measured by ellipsometry over time. This experiment is repeated for a range of humidity values between 50 and 95\%. The measured thicknesses are converted to swelling ratios by dividing them by the dry height.

The (equilibrium) swelling ratios exhibit a strong superlinear relation with the humidity, in accordance with previous studies \cite{ritsema_van_eck_vapor_2022}; at low humidity (<60\%), there is barely any swelling, while above that, small changes in humidity result in large changes in swelling. For these higher humidities, an equilibrium is not fully reached within the \SI{90}{\second} timeframe of the measurement.

The swelling results suggest the absorption dynamics are slower to reach equilibrium for higher humidities, while desorption is relatively quicker for brushes swollen at higher humidities. Furthermore, for low humidity values, the swelling curves show noticeable kinks, which could indicate anomalous diffusion in the brush. This is seen more clearly in the inset in \autoref{fig:elip}.

For simple diffusion polymer swelling systems, the Berens-Hopfenberg model \cite{berensDiffusionRelaxationGlassy1978} (\autoref{eq:BH}) is often used to describe the swelling characteristics. The model describes the swelling characteristics as a sum of a diffusive component and a chain relaxation component. In this equation, $S_{\rm T}$, $S_{\rm F}$, and $S_{\rm R}$ represent the total, diffusive, and relaxation components of the swelling, $D$ is the diffusion coefficient of the solvent in the polymer matrix, $t$ is the time, $L$ is the thickness of the brush, and $\tau_{\rm R}$ is the relaxation time constant of the polymer matrix \cite{crankMathematicsDiffusion1979, ogiegloTemperatureinducedTransitionDiffusion2013, visser_when_2007}.

\begin{align}
    S_{\rm T} = &S_{F, \inf} \left[1 - \frac{8}{\pi^2}\sum^{\inf}_{m=0}\frac{\exp\left(-\frac{D\left(2m+1\right)^2\pi^2t}{L^2}\right)}{\left(2m+1\right)^2} \right] \nonumber\\
    + &S_{R, \inf} \left[1 - \exp\left(\frac{t}{\tau_{\rm R}}\right)\right]
    \label{eq:BH}
\end{align}

A fit of the Berens-Hopfenberg model to the swelling curves in \autoref{fig:elip} is shown in SI-1, but does not adequately fit the data. As more complex transport models risk over-describing the data, we choose not to fit our data to a model.

The timescale for swelling and the levelling off of the signal is much shorter for infrared spectroscopy compared to ellipsometry. This is explained by the nature of the measurement. In infrared spectroscopy, not the water absorption but the change of absorption of the brush is measured. This shift happens before the brush is saturated because this shift is dependent on the interaction of the brush with the absorbed water. Once brush-water interaction is saturated, the brush continues to swell because this is energetically favourable. However, the swelling no longer alters the brush absorption, as the studied bonds are already fully engaged in water-brush interaction.

\section{Conclusion and outlook}
We have shown the application of a home-built high sampling rate infrared spectrometer in the study of a highly reproducible chemical system. The transients of brush dynamics using infrared spectroscopy occur at a different timescale than the total brush swelling, as is shown from the ellipsometry reference experiment. While this discrepancy is phenomenologically explained, further inference on the physics of solvent-induced polymer brush swelling dynamics from both ellipsometry and infrared spectroscopy data requires a transport model. While more elaborate models are available in the literature, these were not used because they might over\-describe the data. The spectrometer setup, and DF-IR, is highly suitable towards the study of a single transient, or the study of a spectrum for a highly reproducible system.

\section*{Acknowledgements}
The authors would like to thank Jasper Lozeman and Matthias Godejohann for their insightful input on infrared spectroscopy. Furthermore, we thank Joshua Loessberg-Zahl, Nieck Benes, and Guido Ritsema van Eck for discussions on transport theory in thin polymer layers. 

\emergencystretch=1em
\printbibliography

\end{document}